\documentclass[aps,prb,twocolumn,showpacs,psfig,superscriptaddress,longbibliography]{revtex4-1}

\usepackage{textcomp}
\usepackage{times}
\usepackage{graphicx}
\usepackage{float}
\usepackage{latexsym,amsmath,amssymb,bm,euscript}
\usepackage{color}
\usepackage{subfigure}
\usepackage{epstopdf}
\usepackage[colorlinks=true,linkcolor=blue,citecolor=blue]{hyperref}
\usepackage{hyperref}
\usepackage{soul}
\usepackage[normalem]{ulem}
\usepackage{mathrsfs}
\usepackage{amsmath}
\usepackage{lettrine}
\usepackage{xspace}
\usepackage{textcomp}

\begin{document}

\title{Pressure-induced superconductivity in flat-band Kagome compounds Pd$_3$P$_2$(S$_{1-x}$Se$_x$)$_8$ }

\author{Shuo Li}
\thanks{These authors contributed equally to this study.}
\affiliation{Mathematics and Physics Department, North China Electric Power
University, Beijing 102206, China}
\affiliation{Department of Physics and Beijing Key Laboratory of
Opto-electronic Functional Materials $\&$ Micro-nano Devices, Renmin
University of China, Beijing, 100872, China}

\author{Shuo Han}
\thanks{These authors contributed equally to this study.}
\affiliation{Department of Physics and Beijing Key Laboratory of
	Opto-electronic Functional Materials $\&$ Micro-nano Devices, Renmin
	University of China, Beijing, 100872, China}

\author{Shaohua Yan}
\thanks{These authors contributed equally to this study.}
\affiliation{Department of Physics and Beijing Key Laboratory of
	Opto-electronic Functional Materials $\&$ Micro-nano Devices, Renmin
	University of China, Beijing, 100872, China}

\author{Yi Cui}
\affiliation{Department of Physics and Beijing Key Laboratory of
Opto-electronic Functional Materials $\&$ Micro-nano Devices, Renmin
University of China, Beijing, 100872, China}

\author{Le Wang}
\affiliation{Department of Physics and Beijing Key Laboratory of
Opto-electronic Functional Materials $\&$ Micro-nano Devices, Renmin
University of China, Beijing, 100872, China}

\author{Shanmin Wang}
\affiliation{Department of Physics, Southern University of
Science $\&$ Technology, Shenzhen, Guangdong, 518055, China}

\author{Shanshan Chen}
\affiliation{Department of Physics and Beijing Key Laboratory of
Opto-electronic Functional Materials $\&$ Micro-nano Devices, Renmin
University of China, Beijing, 100872, China}

\author{Hechang Lei}
\email{hlei@ruc.edu.cn}
\affiliation{Department of Physics and Beijing Key Laboratory of
	Opto-electronic Functional Materials $\&$ Micro-nano Devices, Renmin
	University of China, Beijing, 100872, China}

\author{Feng Yuan}
\email{yuan@qdu.edu.cn}
\affiliation{College of Physics, Qingdao University, Qingdao 266071, China}

\author{Jinshan Zhang}
\email{zhangjs@ncepu.edu.cn}
\affiliation{Mathematics and Physics Department, North China Electric Power
University, Beijing 102206, China}

\author{Weiqiang Yu}
\email{wqyu\_phy@ruc.edu.cn}
\affiliation{Department of Physics and Beijing Key Laboratory of
Opto-electronic Functional Materials $\&$ Micro-nano Devices, Renmin
University of China, Beijing, 100872, China}


\begin{abstract}

We performed high-pressure transport studies on the flat-band Kagome compounds,
Pd$_3$P$_2$(S$_{1-x}$Se$_x$)$_8$ ($x$~=~0, 0.25), with a diamond anvil cell.
For both compounds, the resistivity exhibits
an insulating behavior with pressure up
to 17 GPa. With pressure above 20~GPa, a metallic behavior is observed at high
temperatures in Pd$_3$P$_2$S$_8$, and superconductivity emerges at low temperatures.
The onset temperature of superconducting transition $T_{\rm C}$ rises monotonically from 2~K to 4.8~K
and does not saturate with pressure up to 43 GPa. For the Se-doped compound
Pd$_3$P$_2$(S$_{0.75}$Se$_{0.25}$)$_8$,
the $T_{\rm C}$ is about 1.5~K higher than that of the undoped one over the whole
pressure range, and reaches  6.4~K at 43~GPa.
The upper critical field with field applied along the $c$ axis at typical pressures is
about 50$\%$ of the Pauli limit, suggesting a 3D superconductivity.
The Hall coefficient in the metallic phase is low and exhibits a peaked
behavior at about 30~K, which suggests either a multi-band electronic
structure or an electron correlation effect in the system.

\end{abstract}

\maketitle


Kagome lattice systems have been proposed to host rich physics, such as quantum spin liquid
and unconventional superconductivity in systems with strong electron correlations~\cite{1987_Anderson_Science, 2009_Wen_PRB,2012_Lee_Nature,2019_LiJX_QuantumM,2021_SLLi_CPL}.
Recently, the electronic flat band~\cite{2018_Zhang_PRL,2019_hasan_NP,2020_Comin_NM,2020_Comin_NC,2020_WangSC_NC,2021_WangSC_NC,2021_Checkelsky_NC,2021_Checkelsky_arXiv,2020_Park_SR,2021_LeiHC_PRB}
and Dirac cones~\cite{2020_Comin_NC,2020_WangSC_NC,2021_WangSC_NC,2018_Claudia_NaturePhy,2018_Joseph_Nature,2021_LeiHC_CPL,2021_JGCheng_PRL,2022_XHChen_Nature,2021_JPHu_ScienceBulletin,2021_GaoHJ_CPL} were also observed and caused intense research interests
in several metallic kagome materials, where the electron correlations appear to be weak.
In the Kagome lattice, the flat band arises from phase destruction
of electron wave functions in the corner shared triangles, so that one electron band is localized
in the hexagons~\cite{1991_Mielke_JoPA3,2013_Thomale_PRL}.
Theoretically, the high electronic density of states in the flat band
is expected to enhance the correlation effect if the flat band is close to the Fermi surface,
which may give rise to unusual phenomena such as high $T_{\rm C}$ superconductivity\cite{2000_kohno_PRL,2007_Furukawa_PhysicsC,2009_Wen_PRB},
fractional quantum Hall effect\cite{2011_Wen_PRL} and ferromagnetism\cite{1991_Mielke_JoPA2,1998_Tasaki_TheoreticalPhy}.

Unfortunately, the electron correlations in the flat-band Kagome materials seem
to be too weak to offer any novel phase so far~\cite{2018_Zhang_PRL,2019_hasan_NP,2020_Comin_NM,2020_Comin_NC,2020_WangSC_NC,2021_Checkelsky_NC,2021_WangSC_NC,2021_Checkelsky_arXiv,2020_Park_SR,2021_LeiHC_PRB}.
In parallel, signatures of flat bands and the related novel phases
have been discovered experimentally in twisted bilayer
graphene~\cite{2018_Pablo_nature,2018_Pablo_Nature_2}, silicene~\cite{2018_Du_SA} and dichalcogenides~\cite{2020_Dean_NM}.
An alternative way to look for novel phases
is to tune the materials by doping or pressure\cite{2020_RenC_CPL,2021_ChenXL_CPL}, so that the
electrons can be partly delocalized and bring in the electron correlation
effect. In this regard, candidate Kagome
materials having flat band close to Fermi surface,
such as CoSn~\cite{2020_Sales_PRB,2020_WangSC_NC,2020_Comin_NC} and Pd$_3$P$_2$S$_8$~\cite{2020_Park_SR,2021_LeiHC_PRB},
are favored for further study by doping or pressure tuning.

\begin{figure}[t]
	\includegraphics[width=8cm]{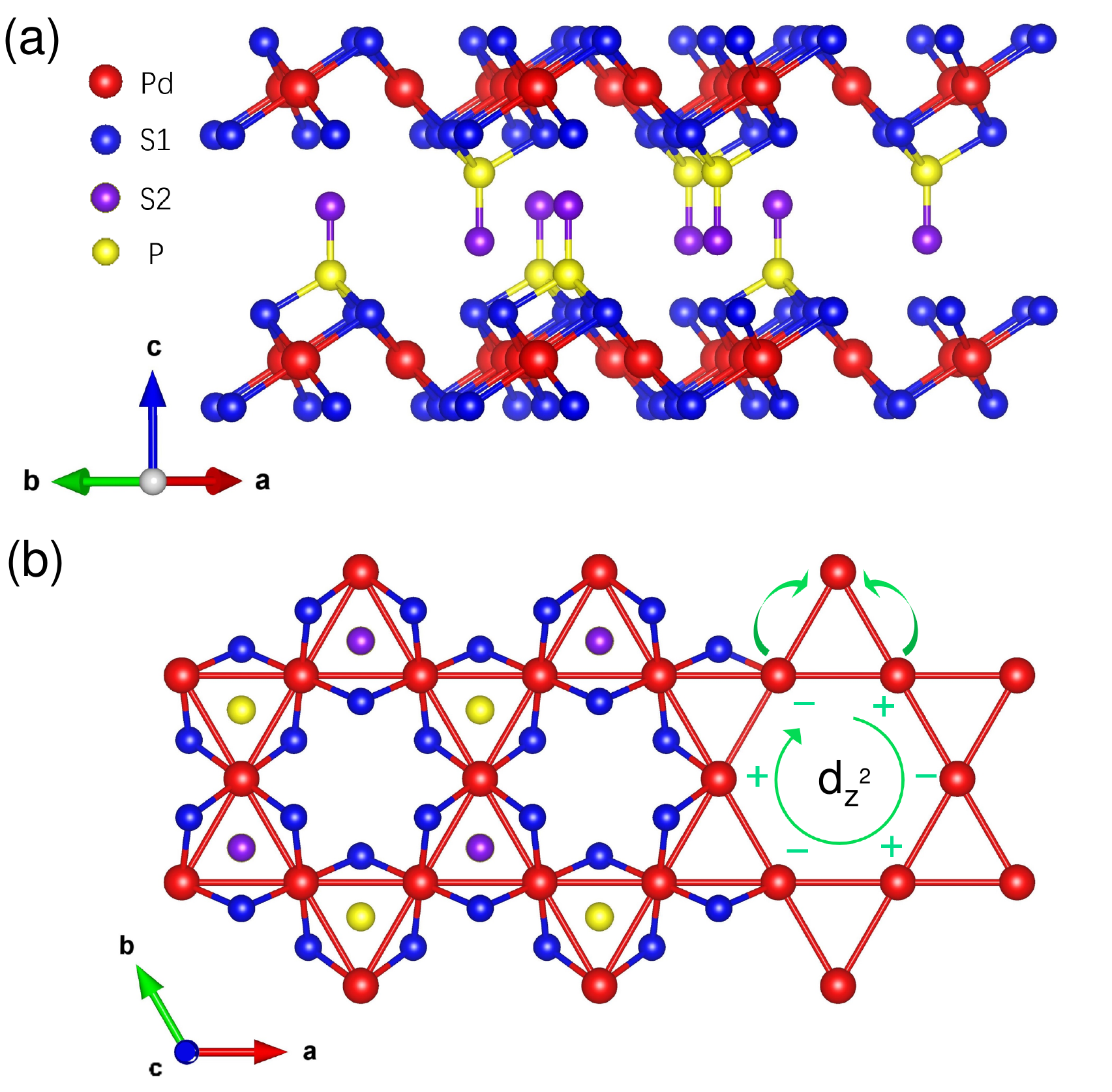}
	\caption{\label{struc}
    (a) Planar view of the crystal structure of Pd$_3$P$_2$S$_8$.
	(b) Top view of one kagome layer. Phase destruction of the electron hopping is
     illustrated in a triangle, and electron localization is present in the hexagon.
     $d_{z^2}$ labels the primary orbital that forms the flat band.
}
\end{figure}

\begin{figure*}[t]
\includegraphics[width=17cm]{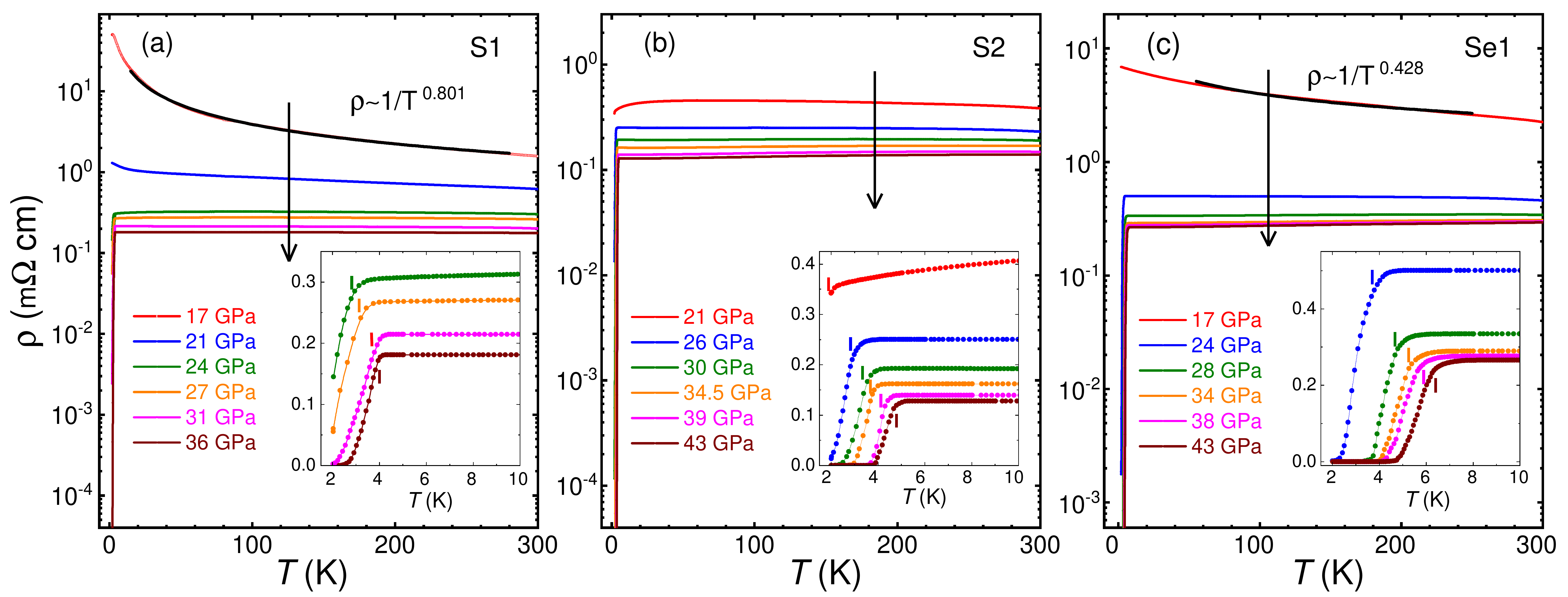}
\caption{
\label{rvst}
  (a) In-plane resistivity as functions of temperatures of undoped sample S1, measured under different pressures.
  The black solid line represents a power-law function fit, ${\rho} {\sim} 1/T^{0.801}$, to the data at 17~GPa.
  Insert: An enlarged view of low-temperature resistivity.
  (b)-(c) In-plane resistivity of undoped sample S2 and the Se-doped sample Se1, respectively, and an enlarged
  view of low-temperature resistivity in the inset. The black solid line in (c) is a function fit of ${\rho} {\sim} 1/T^{0.428}$ to the data at 17~GPa.
  Short vertical lines in the insets of (a)-(c) mark the 10$\%$ drop of the normal-state resistivity  which signals
  the onset of superconductivity.
}
\end{figure*}

Here we report our high-pressure transport studies on a van-der-Waals Kagome
compound Pd$_3$P$_2$S$_8$ and its Se-doped compound Pd$_3$P$_2$(S$_{0.75}$Se$_{0.25}$)$_8$,
both of which are highly insulating at the ambient pressure~\cite{2020_Park_SR,2021_LeiHC_PRB}.
Pd atoms in Pd$_3$P$_2$S$_8$ constitute the Kagome plane, as shown
in Fig.~\ref{struc}. Local density approximation (LDA) calculations on Pd$_3$P$_2$S$_8$
reveal that a flat band, mainly formed by the $d_{z^2}$  orbital of Pd$^{2+}$,
is located at about 0.2~eV below the Fermi surface~\cite{2021_LeiHC_PRB}.
Se doping enhances the interlayer coupling and reduces the gap of the flat band~\cite{2021_LeiHC_PRB}.

Our high-pressure transport studies show that for both compounds, the conductivity at 300~K
is enhanced under pressure. At about 17~GPa, an insulating behavior
is still seen. However, their conductivity exhibits a power-law temperature dependence
in two decades of temperature below 300~K, which suggests strong electron localization
caused by disorder~\cite{1977_Thouless_PRL} .
With further increase of pressure above 20~GPa, a metallic behavior is induced
at high temperatures. Superconductivity emerges at low temperatures
in the metallic phase. The upper critical field is also found very high with field applied
along the $c$ axis.
A small Hall coefficient is also observed in the metallic phase and exhibits a strong
temperature dependence, which suggests that the system
may have an electronic correlation effect.

Pd$_3$P$_2$(S$_{1-x}$Se$_x$)$_8$  ($x$=0, 0.25) single crystals were
grown by the chemical vapor transport method~\cite{2020_Park_SR,2021_LeiHC_PRB}. 
As shown in Fig.~\ref{struc}(a), Pd$_3$P$_2$S$_8$  has a 
quasi-2D structure with S1-Pd-S1-P-S2 stacked sequentially along the
$c$ axis with an interlayer distance $c$~=~7.247~$\rm \AA$~\cite{2021_LeiHC_PRB}, 
and Pd atoms form a perfect kagome structure with the 
atomic distance $d_{\rm {Pd}\mbox{-} \rm {Pd}}$~=~3.418~$\rm \AA$~\cite{1971_YoungHS_JoSSC}.
Se doping replaces S atoms randomly and enhances the interlayer coupling~\cite{2021_LeiHC_PRB},
which leads to a larger dispersion and a reduction of gap of the flat band.
Note that $x$~=~0.25 is the maximal Se doping level currently available~\cite{2021_LeiHC_PRB}.

We performed high-pressure transport measurements with a BeCu diamond anvil cell (DAC).
Diamond anvils of 300~$\mu$m culets and a rhenium gasket covered with cubic boron nitride (c-BN)
powder were used, with no pressure transmitting medium filled in.
The highest  pressure we achieved is 43~GPa.
The pressures were calibrated by the Raman spectrum from the culet of a top
diamond anvil and fluorescence from the ruby using a Raman microscope (WITec Alpha 300R).
For the transport measurements, the DAC, with sample loaded inside, is cooled in a Quantum Design
Physical Property Measurement System (PPMS-14T). The in-plane resistivity and the Hall resistivity
of samples were performed by the standard van der Pauw method.
Three single crystals, cut into dimensions of 100$\times$100$\times$15 $\mu$m$^3$, were measured.
Two of them are undoped, labeled as S1 and S2 respectively, and one is doped, labeled as Se1.


The resistivity of three crystals is first measured with increasing pressures at zero field,
and cooled from 300~K down to 2~K.
Detailed data are shown as functions of temperatures in Fig.~\ref{rvst}(a)-(c).
At ambient pressure, the resistance $\rho$ of all three samples
is large at room temperature and is barely measurable at low temperatures.
This is consistent with the gapped behavior,
where all the bands are away from the Fermi surface~\cite{2021_LeiHC_PRB}.
With applied pressure, the resistivity $\rho$ decreases
for all samples, as shown in Fig.~\ref{rvst}(a)-(c).
For sample S1, the insulating behavior still holds under pressures of 17 and 21~GPa,
as shown by the increase of resistivity upon cooling.

At 17~GPa, the conductivity of sample S1 is found to follow a power-law temperature dependence,
that is, ${\rho} {\sim} 1/T^{\alpha}$ with the power-law exponent $\alpha=$~0.801$\pm$0.001,
as shown by the fit in Fig.~\ref{rvst}(a). The fitting holds in nearly two orders of
temperature range below 300~K. By contrast, our data cannot be well fit to a gap function.
This power-law temperature dependence of conductivity suggests that insulating behavior
is caused by strong electron localization, as a result of disorder
in the system~\cite{1977_Thouless_PRL,2002_Efros_PRL}.
In fact, it is found that $\alpha$~=~1 for two-dimensional (2D)
systems~\cite{1985_Imry_JoPc} and
$\alpha$ ranges between 1/3 and 1/2 for 3D systems~\cite{1997_Witcomb_JoP,2002_Efros_PRL}.
Our value of $\alpha\approx$~0.8 indicates that the system
crossovers from the 2D to 3D at 17 GPa, which suggests that
interlayer coupling is strongly enhanced under pressure.
Note that the sample at the ambient pressure shows a high quality as revealed by
our XRD data~\cite{2021_LeiHC_PRB}, which leads us to speculate that the disorder effect
is either an intrinsic pressure-induced structural disorder of the material, or
brought in by pressure inhomogeneity under such high pressures.

A very weak upturn in resistivity is still observed at 21~GPa for sample S1,
which suggests that the insulating behavior tends to be suppressed at pressures above 20~GPa.
At and above 23~GPa, the resistivity decreases weakly upon cooling, which is a clear signature
of a metallic behavior.
An enlarged view of the low-temperature resistivity,
at pressures above 20~GPa, is further demonstrated in the
inset of Fig.~\ref{rvst}(a). A high conductivity is seen
with resistivity less than 0.5~m${\Omega}$~cm at room temperature.
Since the resistivity does not show a large upturn upon cooling at low temperatures,
a metallic, rather than an insulating behavior, is clearly established
with pressure above 20~GPa.

\begin{figure}[t]
	\includegraphics[width=8cm]{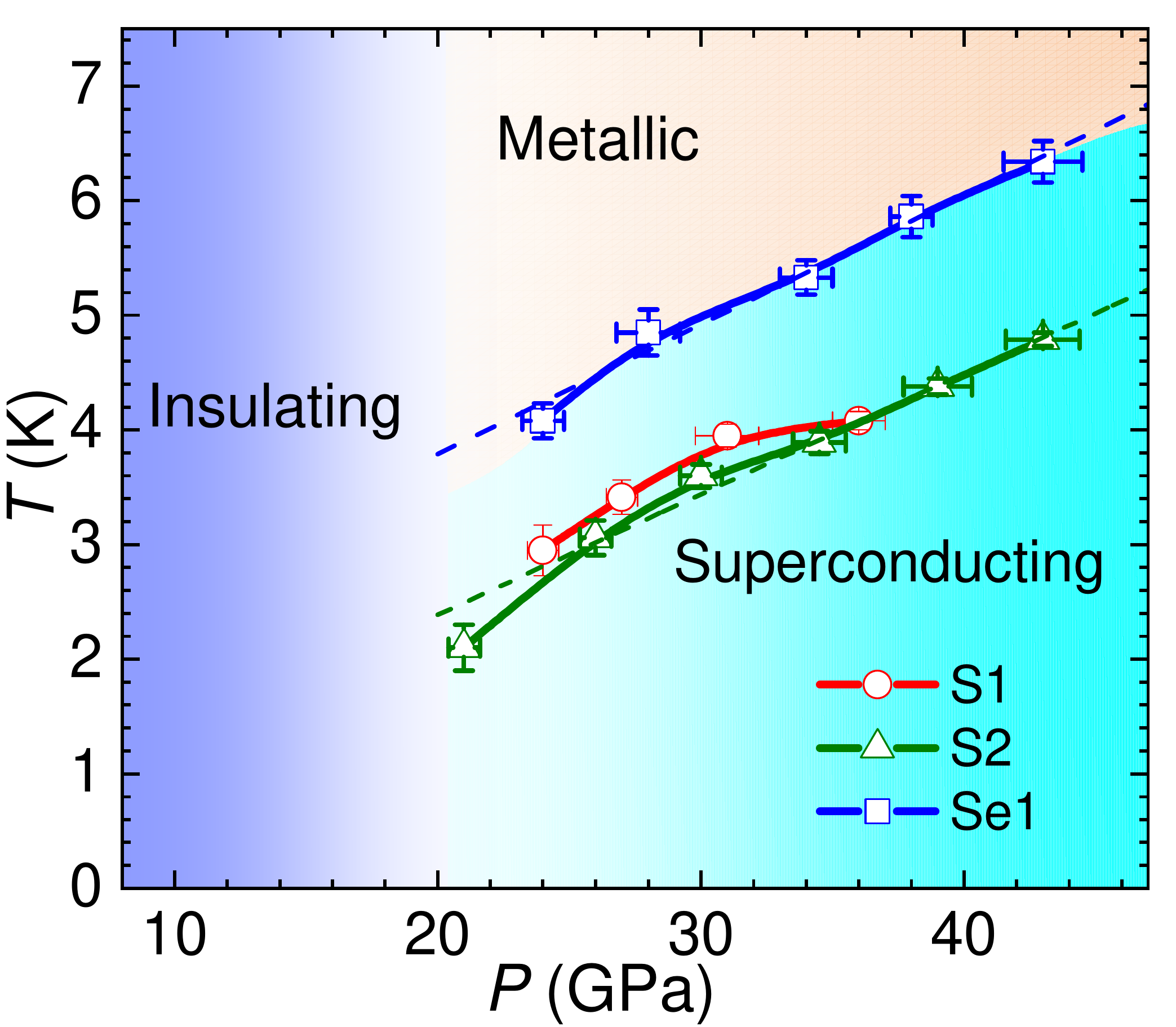}
	\caption{\label{pd}
	The high-pressure phase diagram established on samples S1, S2, and Se1.
    The insulating (metallic) phase labels the regime where a prominent resistivity upturn is observed (absent) upon cooling. Open circles,
	triangles, and squares represent the onset $T_{\rm C}$  of the three samples, with the superconducting phase located below.
    The straight dashed lines are linear fit for the undoped and the Se-doped samples with the slope 0.105~K/GPa and 0.113~K/GPa, respectively.
    }
\end{figure}

Therefore, an insulating to metallic transition is revealed at
about 20~GPa in samples S1. Similar transition is also observed for sample S2
and Se1, as shown in Fig.~\ref{rvst}(b)-(c). Interestingly, the
insulating to metallic transition for all three samples
occurs at pressures close to 20~GPa, as shown by the
disappearance of the resistivity upturn in Fig.~\ref{rvst}(b)-(c).
For sample Se1, the conductivity at 17~GPa also follows the power-law fit with $\alpha\approx$~0.428 in a limited temperature range, as shown in Fig.~\ref{rvst}(c).
Such a lower $\alpha$ suggests that the doped sample has a strong 3D feature at the same pressure~\cite{1997_Witcomb_JoP,2002_Efros_PRL},
which is consistent with the LDA calculation that Se doping
enhances interlayer coupling~\cite{2021_LeiHC_PRB}.

When the metallic behavior is established at high temperatures,
a sudden drop of resistivity for all three samples is observed when further cooled,
as shown in the inset of Fig.~\ref{rvst}(a)-(c).
At even lower temperatures, zero resistivity is reached, which is clear evidence
of superconductivity. The superconducting transition temperature, labeled as $T_{\rm C}$, is defined by
the temperature at which a 10$\%$ drop of normal-state resistivity is achieved, as shown
in the inset of Fig.~\ref{rvst}(a)-(c). It is noticed that $T_{\rm C}$
rises continuously with pressure up to 43~GPa.


The $T_{\rm C}$ is then determined and plotted as functions
of pressures for three samples in the phase diagram in Fig.~\ref{pd}.
Below 20~GPa, an insulating phase is established by
the prominent resistivity upturn; whereas above 20~GPa, the absence of resistivity
upturn reveals an onset of a metallic phase which turns into a superconductor
at low temperatures.
$T_{\rm C}$ rises monotonically
with pressure for both the undoped and the doped crystals. For the measured pressure range,
$T_{\rm C}$ increases nearly linearly with pressure,
at least up to 43~GPa.
The slope is determined as $dT_{\rm C}/dP$~=~$0.11\pm0.05$~K/GPa for all samples.
At pressures above 30~GPa, a small increase in $dT_{\rm C}/dT$ is also observed,
which suggests that a higher $T_{\rm C}$ should be achieved at pressures
above 43~GPa.

It is also interesting to note that the $T_{\rm C}$ for the doped sample
is about 1.5~K higher than the undoped one in the whole pressure range.
In all compounds, superconductivity is not observed with pressure below 20~GPa,
regardless of doping. This will be further discussed later.

\begin{figure}[t]
	\includegraphics[width=8.5cm]{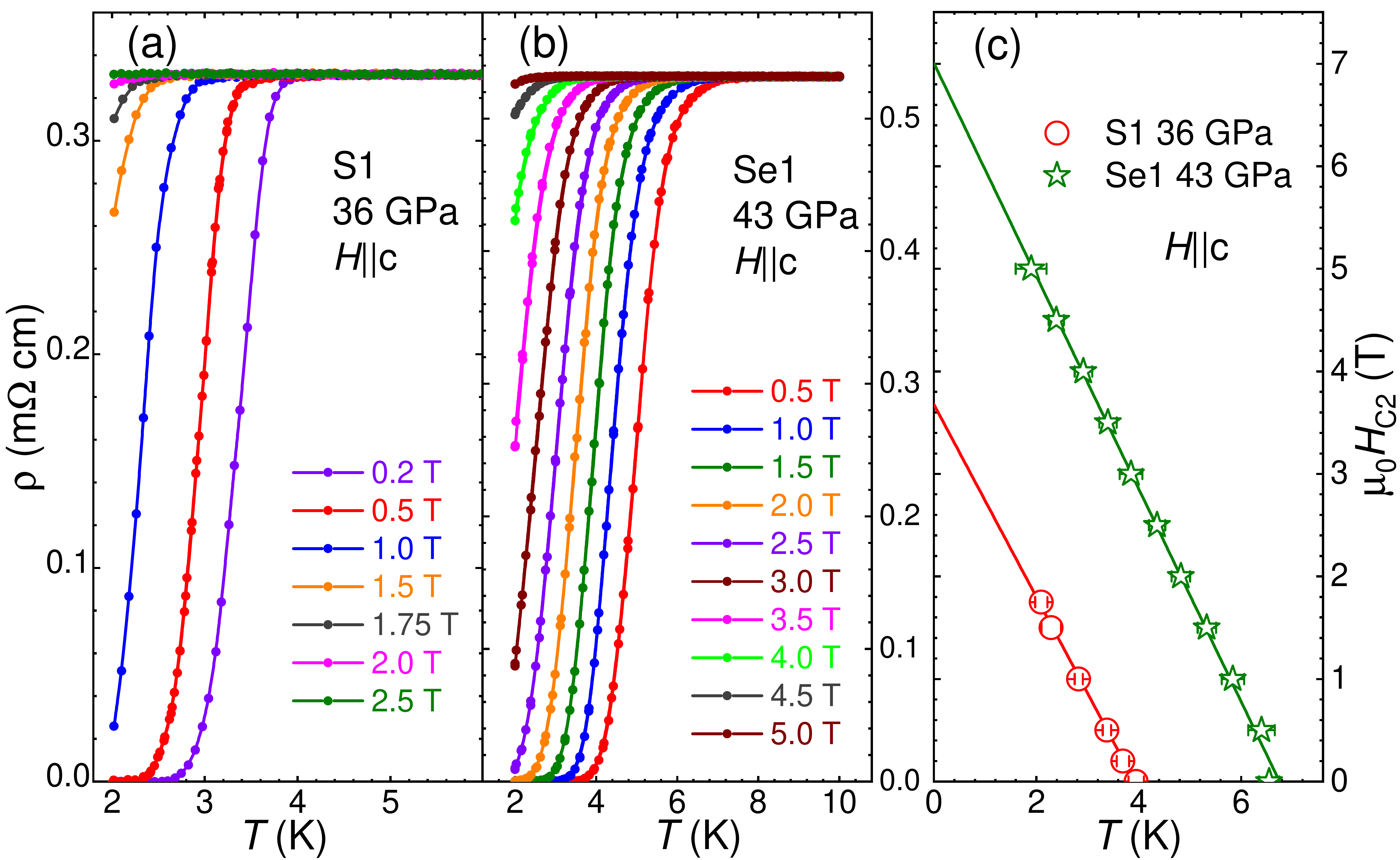}
	\caption{\label{rvsh}
	(a)-(b) Resistivity of samples S1 and Se1, as functions of temperatures measured
     with fields applied along the $c$ axis, under pressure of 36~GPa and 43~GPa, respectively.
     (c) The out-of-plane $H_{\rm C2}$ of samples S1 and Se1 as functions of temperatures. The $H_{\rm C2}$ is defined by the magnetic field where the normal-state resistivity drops by 10$\%$ shown in (a) and (b). The solid lines in (c) are linear function fits of $H_{\rm C2}(T)$ for samples S1 and Se1, respectively.
    }
\end{figure}

Recently, two other studies~\cite{2022_ZRYang_arXiv,2022_YPQi_arXiv} on Pd$_2$P$_2$S$_8$ also report the pressure-induced superconductivity after this work, with $T_{\rm C}$ increasing monotonically with pressure.
In Ref.~\citenum{2022_ZRYang_arXiv}, the onset pressure for superconductivity is about 10~GPa lower than ours, and $T_{\rm C}$ is about 2~K higher at the same pressures.
In fact, a pressure-induced amorphous phase is found by the XRD and Raman scattering, which emerges simultaneously with superconductivity~\cite{2022_ZRYang_arXiv,2022_YPQi_arXiv}.
The difference of $T_{\rm C}$ may be caused by the pressure transmission medium used in different groups, since the amorphous phase should rely sensitively
on the pressure conditions, as revealed by the strong hysteresis of superconductivity with pressure~\cite{2022_ZRYang_arXiv}.

The out-of-plane upper critical field of superconductivity is also
studied by the resistivity measurements with field applied along the crystalline $c$ axis.
The resistivity of samples S1 and Se1 are shown as functions of temperatures
in Fig.~\ref{rvsh}(a)-(b) with different fields, each under a fixed pressure.
For each fixed field, the resistivity of sample S1 measured at 36 GPa, as shown
in Fig.~\ref{rvsh}(a), first rises from zero with increasing temperature.
Above a specific temperature, resistivity exhibits a level off, which
signals the disappearance of superconductivity;
then the upper critical field $H_{\rm {C2}}$ at that
temperature is determined as shown in Fig.~\ref{rvsh}(a).
The $H_{\rm C2}$ of sample Se1 is also determined in
the same way with data shown in Fig.~\ref{rvsh}(b), under a
fixed pressure of 43 GPa.

The $H_{\rm C2}$ of both crystals is plotted as functions of temperatures
in Fig.~\ref{rvsh}(c), which decreases with temperature. At low temperatures, $H_{\rm{C2}}$ drops almost linearly with field.
Extrapolation to the zero temperature limit
gives $H_{\rm C2}$($T$=0)~$\approx$~3.68~T for sample S1, and
$H_{\rm C2}$(0)~$\approx$~7.0~T for sample Se1.
Considering the onset $T_ {\rm C}$ at zero field is about 4.08~K (6.4~K) for
sample S1 (Se1), the zero temperature
$H_{\rm C2}$ is then about $50\%$ of the Pauli limit~\cite{1996_Gonzalez_PRB,1973_Fulde_Advance,2019_Takeyama_Scireport}, assuming $g$~=~2.
Such a high out-of-plane $H_{\rm C2}$ suggests that the system is likely a 3D
superconductor, which is consistent with the resistivity measurements
where a 2D to 3D crossover is already observed at 17~GPa as discussed above.
Note that a high out-of-plane critical field is also reported in the
iron-based superconductors~\cite{2009_Yuan_Nature}.


\begin{figure}[t]
\includegraphics[width=8.5cm]{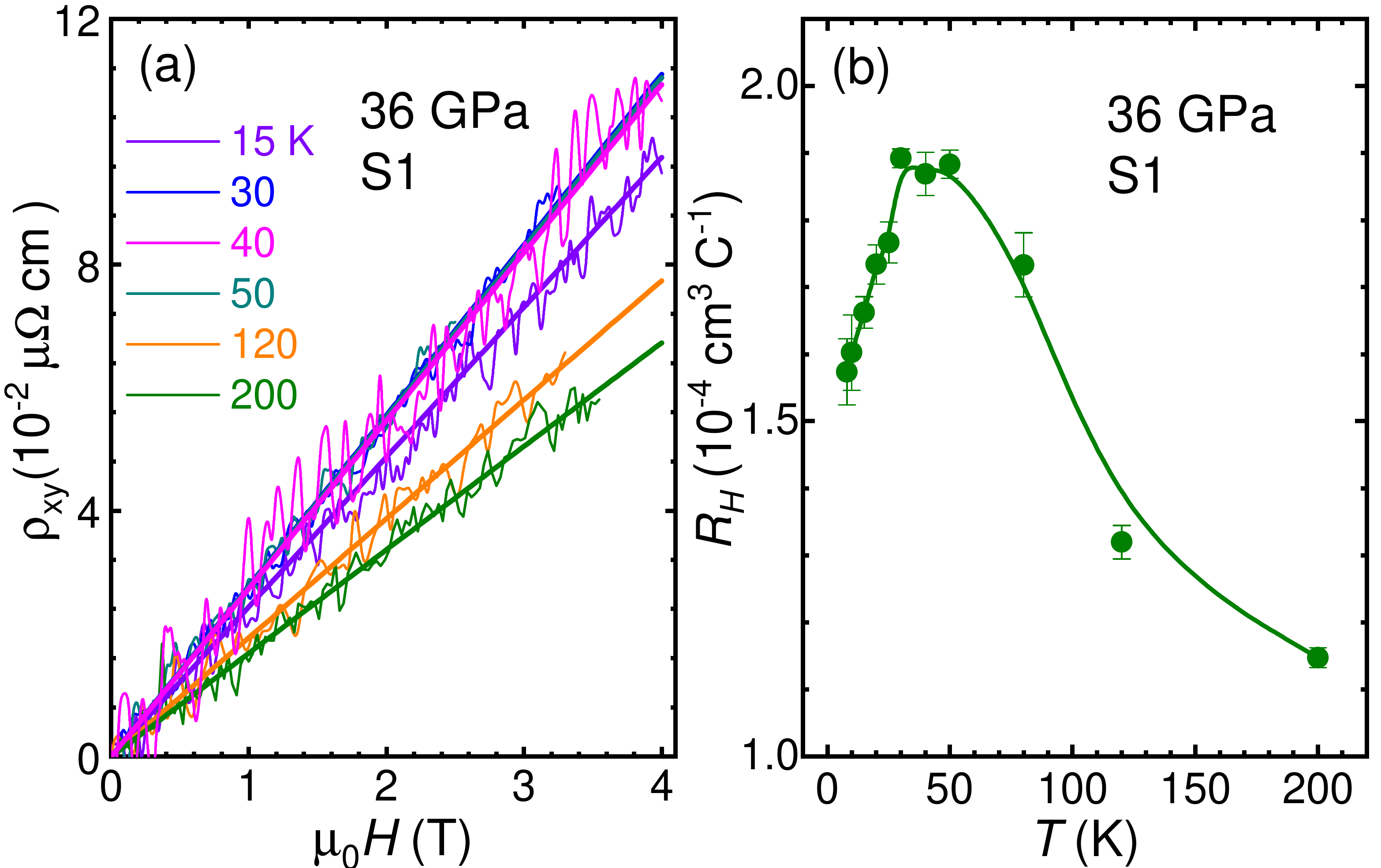}
\caption{\label{hall}
	(a) The in-plane Hall resistivity $\rho_{xy}$ of sample S1 as functions of fields applied along the $ c$ axis, measured at
       different temperatures under a constant pressure 36~GPa. The straight lines
       are linear function fits to the data to calculate the Hall coefficient $R_{\rm H}$.
    (b) The obtained $R_{\rm H}$ as a function of temperature.
 }
\end{figure}

To further understand the carrier properties, the Hall resistivity $\rho_{xy}$
on sample S1 is measured with field applied along the $c$ axis, under a pressure of 36~GPa.
As shown in Fig.~\ref{hall}(a), $\rho_{xy}$, plotted as functions of
fields measured at different temperatures, shows a linear
field dependence with field up to 4~T.
By taking the slope of the Hall
resistivity, the Hall coefficient $R_{\rm H}$ is then calculated and plotted
as a function of temperatures in Fig.~\ref{hall}(b).
Both $\rho_{xy}$ and $R_{\rm H}$ are calibrated by the high-pressure
lattice parameters reported in Ref.~\citenum{2022_YPQi_arXiv}. $R_{\rm H}$ stays
positive with temperatures from 200~K to 8~K, which suggests
that the major carriers are the hole type in
the system under pressure. However, $R_{\rm H}$ changes largely
with temperature and is peaked at about 30~K.
Such a dramatic temperature dependence  suggests
that the compounds either have multi electronic bands on the Fermi surface~\cite{1976_Mermin_book,2007_Greene_PRL},
and/or have strong electron correlations~\cite{1994_Peck_PRL,2007_Greene_PRL}.

In fact, the values of $R_{\rm H}$ in the whole temperature range are very small,
whose peaked value reads 1.89 ${\times}$ 10$^{-4}$ cm$^3$C$^{-1}$ at 30~K.
Taking the value of $R_{\rm H}$ at 10~K, the carrier concentration is estimated
as 3.90 $\times$ 10$^{22}$ cm$^{-3}$ at this pressure, assuming a single band. However,
this would suggest that each unit cell contains an excessive carrier density of 9.26 holes per unit cell at 10~K.
To understand the small and temperature-dependent
$R_{\rm H}$, we propose two alternative scenarios as described below.

The first scenario is based on the nearly compensated electron and hole bands
on the Fermi surface. If the system contains both
hole and electron types of carriers at high pressures, the value of $R_{\rm H}$ can be
very small as observed. Together with the large
temperature-dependent $R_{\rm H}$ as described above, a pressure induced
multi-band system, with both electron and hole types of carriers on the
Fermi surface, is suggested for the high pressure phase.
Comparing with the band structure calculations~\cite{2021_LeiHC_PRB},
the d$_{z^2}$ band of Pd$^{2+}$ is only slightly below the Fermi surface,
which may constitute the hole band.
By contrast, the d$_{x^2-y^2}$ orbital is located at
about 1.5~eV above the Fermi surface, which may become the electron
band at high pressures.

The second scenario is based on the correlation effect of the flat band.
With more portion of the (nearly) flat band pushed to the Fermi surface
under pressure, both the carrier density
and the electron correlation effect may be strongly enhanced
simultaneously, which result in the temperature-dependent Hall coefficient with a small value.
Given the non-monotonic change of the Hall coefficient with
temperature, the electron correlation scenario is favored.

To our knowledge, this is a rare superconductor derived from
a Kagome compound with a flat band.
We now attempt to understand the pressure and Se doping
effect in all the samples.
As reported, the emergence of superconductivity depends on the pressure-induced structural
transition from the crystalline to amorphous phase~\cite{2022_ZRYang_arXiv,2022_YPQi_arXiv}.
The emergence of the amorphous phase is consistent with our resistivity data,
where large disorder is already observed in the low-pressure phase by
the power-law temperature dependence of conductivity. At pressures above 20~GPa,
however, the large carrier density may mask the power-law behavior in the conductivity.
In principle, further increase of pressure may push the flat band more dispersive,
which results in higher carrier concentration on the Fermi surface and enhances
 $T_{\rm C}$  consequently. Se doping may also enhance  $T_{\rm C}$  by the increase of
the hoping elements with more extended orbitals of Se.
On the other hand, the emergence of superconductivity relies only on the external pressure
which causes the amorphous phase, so that superconductivity occurs
at a similar pressure for both the parent and the doped compounds.



In summary, our high-pressure transport studies on the undoped and the Se-doped
Pd$_3$P$_2$S$_8$ reveal that the resistivity changes from
an insulating to a metallic behavior at about 20~GPa.
The power-law temperature dependence of conductivity at low pressure suggests
a high level of disorder induced by pressure, with a 2D to 3D crossover.
Superconductivity is observed in the metallic phase for all the compounds.
The out-of-plane $H_{\rm C2}$ is high, which again suggests a 3D nature of
superconductivity. The superconducting transition
temperature $T_{\rm C}$ rises with pressure at least up to 43~GPa.
$25\%$ Se doping further enhances $T_{\rm C}$ by 1.5~K.
Our data suggest that the high-pressure phase likely has
both electron and hole bands on the Fermi surface, or strong electron correlations,
as revealed by the small value of the Hall
coefficient and its large, non-monotonic temperature dependence.

W.Y. and Y.C. are supported by the National Natural Science Foundation of China
under Grants No.~51872328,~12134020 and~12104503, the
Ministry of Science and Technology of China under Grant No.~2016YFA0300504,
the Fundamental Research Funds for the Central
Universities and the Research Funds of Renmin University of China
under Grants No.~21XNLG18 and No.~18XNLG24.
Y.C. is also supported by the China Postdoctoral Science Foundation under Grant
No.~2020M680797.
H.C.L is supported by the National Key R\&D Program of China under Grant No. 2018YFE0202600, the Beijing Natural Science Foundation under Grant No. Z200005, the Fundamental Research Funds for the Central Universities and the Research Funds of Renmin University of China under Grants No. 18XNLG14, 19XNLG13, and 19XNLG17.

\bibliography{Pd3P2S8}

\end{document}